# Prediction of a new efficient permanent magnet SmCoNiFe$_3$

P. Söderlind[1], A. Landa[1], I. L. M. Locht[2], D. Åberg[1], Y. Kvashnin[2], M. Pereiro[2], M. Däne[1], P. E. A. Turchi[1], V. P. Antropov[3], and O. Eriksson[2]

[1]Lawrence Livermore National Laboratory, Livermore, California 94551, USA
[2]Uppsala University, Uppsala, SE-751 20, Sweden
[3]Ames Laboratory, Ames, Iowa 50011, USA

**We propose a new efficient permanent magnet, SmCoNiFe$_3$, that is a breakthrough development of the well-known SmCo$_5$ prototype. More modern neodymium magnets of the Nd-Fe-B type have an advantage over SmCo$_5$ because of their greater maximum energy products due to their iron-rich stoichiometry. Our new magnet, however, removes most of this disadvantage of SmCo$_5$ while preserving its superior high-temperature efficiency over the neodymium magnets.**

Substituting all cobalt atoms for iron, that has a larger magnetic moment, to optimize the maximum energy product (i.e., SmCo$_5$ → SmFe$_5$) makes the ordinary hexagonal phase thermodynamically unstable. This phase, however, is critical for the materials properties and it must be retained for a practical magnet. It is a CaCu$_5$ (D2$_d$)-type structure with three distinct atoms [1] displayed in the lower inset of Fig. 1; The samarium atom (gold, Sm$_1$, Wyckoff position 1a) is centered in a plane with two cobalt atoms surrounding it (dark-blue, Co$_1$, position 2c) and a second layer with three more (light-blue, Co$_2$, position 3g) for a total of six atoms in the unit cell [2]. Because bonding and energetics are dominated by the interaction between 3$d$-orbital electrons from the transition metal and the 5$d$-orbital electrons from samarium [3], it is reasonable to suspect that the crystal instability of SmFe$_5$ is related to a change in the number of 3$d$ electrons due to the substitution of Co for Fe. This proposed scenario is analogous to the crystal stabilities of the magnetic 3$d$ transition metals that are governed by their number of 3$d$ electrons [4].

Indeed, we find a correlation between the number of transition-metal (TM) 3$d$ electrons and the stability of the hexagonal SmTM$_5$ (D2$_d$) compound (TM = Fe, Co, or Ni), see Fig. 1 where we show results from calculations of formation energies. Notice that increasing the 3$d$ electron count (more Ni) greatly stabilizes the compound. An obvious consequence of this behavior is that one should be able to recover crystal stability of a Sm-Fe alloy by doping with 3$d$ electrons from nickel. We compare the theory with experiments [5, 6] that closely reproduce the calculated behavior. Notice that an extrapolation of the experimental heats of formation suggests that an alloy with at least 7.2 3$d$ electrons per TM atom is stable. This is important because SmCoNiFe$_3$ has about 7.3 3$d$ electrons that is enough to stabilize the magnet. It has also recently been confirmed that the related alloy SmNi$_2$Fe$_3$, that has about 7.5 3$d$ electrons, exist [7] in agreement with the experimental extrapolation outlined in Fig. 1.

In a computational search for a thermodynamically viable compound with the maximum portion of iron atoms we recognize that no more than three iron atoms can be accommodated and those occupy the 3g sites that is the most favorable structural position for Fe [8, 9]. For the other (2c) site we investigate a random mix between cobalt and nickel. In the upper inset of Fig. 1 we show the calculated formation energies for the Sm(Co$_x$Ni$_{1-x}$)$_2$Fe$_3$ alloy and here we find that the smallest amount of nickel required to preserve the desired hexagonal phase is about 50% (x ≈ 0.5). This result confirms the observed behavior of the experimental data, namely that SmCoNiFe$_3$ is stable and can be manufactured. The question, then, arises: does this new SmCoNiFe$_3$ alloy have suitable magnetic properties?

High-performance permanent magnets are required to fulfill three intrinsic technical demands: (i) large maximum energy product, a property that relates to the maximal saturation magnetic moment, (ii) high Curie temperature, and (iii) high uniaxial magnetocrystalline anisotropy energy (MAE). First and foremost, the maximum energy product and the magnetic moment are measures of magnetic strength and SmCoNiFe$_3$ exhibits a tremendous moment of ~10 $\mu_B$ mostly due to the iron atoms that each contributes with ~2.7 $\mu_B$. The SmCoNiFe$_3$ moment is thus substantially greater than for the conventional SmCo$_5$ magnet that has a calculated total moment of ~8 $\mu_B$, see Table 1.

The Curie temperature ($T_c$) is essential as well and must be high for the magnet not to fail under warm conditions. It is obtained from calculated exchange-coupling parameters that are mapped onto a Heisenberg model. When comparing the $T_c$ for SmCoNiFe$_3$ relative to SmCo$_5$, we observe only modest decrease from a mean-field (~5 %, 1103 K and 1158 K) and a Monte Carlo (~15%, 790 K and 929 K) method. Both approaches predict $T_c$ in a good accord with the experimental value for SmCo$_5$ ($T_c$ = 1020 K) that is substantially larger than that of the widely used Nd$_2$Fe$_{14}$B magnet ($T_c$ = 588 K) [10].

The third imperative demand on a new functional material is a high MAE. The MAE is the very small energy difference between phases with spin moments oriented in the easy and hard directions, and it is sensitively dependent on an appropriate representation of the electronic and magnetic structures. For a reliable calculation of this quantity one needs an accurate parameter-free first-principles theory. Below we shall describe how this is done and present our result for the MAE: 4.93 meV/cell.

Presently, the best theoretical description for SmCo$_5$ appears to be that of Grånäs *et al.* [11] who published dynamical mean-field theory (DMFT) calculations that adequately captured electronic spectra and magnetic moments. In this work the correlated and localized samarium 4*f* electrons are treated within the Hubbard I approximation (HIA) [11, 12] and the cobalt 3*d* electrons with the spin-polarized T-matrix fluctuation exchange (STPF) solver. The success of the DMFT for SmCo$_5$ is thanks to its proper description of the electronic structure but the method is impractical for calculating magnetic anisotropy due to fundamental and computational obstacles. Therefore, we introduce an alternative approach that is simpler but, as we shall show, captures the main physics of the DMFT while being more workable. In this approach, referred to as the standard rare-earth model (SRM) [12-14], the 4*f* shell is part of the samarium atomic core and not explicitly hybridizing with any valence states. This assumption makes sense because experimentally the Sm-Co alloy system shows essentially no overlap between samarium 4*f* and cobalt 3*d* states and their respective peak intensities are very far apart (6 eV) [15].

We continue with a discussion motivating why the SRM method is a sensible simplification of the full-fledged DMFT-HIA theory by comparing electronic and magnetic structures. In Fig. 2 we show experimental photoemission spectrum (PES) [15] together with our calculated DMFT-HIA and density-functional-theory SRM (DFT-SRM) results for $SmCo_5$. Both models produce spectra close to zero binding energy (Fermi level, $E_F$) with quantitatively correct shape but a peak slightly shifted relative to the experimental PES. However, only the DMFT-HIA captures the deeper lying $4f$ states about 6 eV below $E_F$. Fortunately, for orbital magnetism [16] and magnetic anisotropy, the states closer to $E_F$ are more important and here the DFT-SRM is at least as good as the DMFT-HIA. The good comparison of the electronic spectra supports the DFT-SRM but it is critical that the method can reproduce good magnetic moments as well.

In Table 1 we relate the calculated magnetic moments with polarized-neutron diffraction experiments [17]. Notice that most magnetic moments are very close between the approaches and there is only a minor difference on samarium resulting from their different descriptions of the $4f$ states. The DMFT and DFT magnetic moments also correlate very favorably with several experimental reports [17-19]. The very good agreement in Table 1 between DMFT-HIA and DFT-SRM results combined with the discussion of the electronic spectra above (Fig. 2) makes us confident that the simpler DFT-SRM is reasonable for investigating magnetic anisotropy in $SmCo_5$-related magnets.

There are some subtle points that need to be made from the results of Table 1. First, in the ground-state configuration, the samarium and cobalt magnetic moments align anti-parallel (different sign) in agreement with experimental observations [20]. In the opposite scenario, when these moments couple parallel, as has been assumed in previous models [21, 22], the energy of the system is considerably higher (about 0.1 eV, not shown) and this excited energy state is thus only metastable and not relevant for $SmCo_5$. Second, the orbital magnetic moment on the $Co_1$ atom is greater than on the $Co_2$ atom for both methods in accordance with experiments [17]. This result is particularly relevant for the MAE of $SmCo_5$ because polarized nuclear-magnetic-resonance measurements have shown that the two Co atoms have opposing effects on the magnetic anisotropy [23]. A correct description of the cobalt orbital moments ($Co_1 > Co_2$) is thus necessary for a realistic prediction of the MAE.

One great asset with the DFT-SRM model is that parameter-free calculations of total energies are straightforward without much computational or practical difficulty. From computed total energies of $SmCo_5$ we optimize the crystal structure and obtain a unit-cell volume of 86.0 Å$^3$ and a bulk modulus of 141.9 GPa that both compare satisfactorily with their room-temperature experimental counterparts of 85.74 Å$^3$ and 138.7 GPa [18, 24]. The theoretical hexagonal axial $c/a$ ratio is somewhat small (0.77) relative to the experimental value of 0.798 [18]. From the calculated total energies one can, importantly, also evaluate the MAE.

In Table 2 we contrast our calculated magnetocrystalline anisotropy energies for $\alpha$-Co, $SmCo_5$, and $SmNi_5$ with available experimental data [10, 25, 26]. The most commonly cited value (17.2 MJm$^{-3}$ = 9.2 meV per unit cell) [10] for $SmCo_5$ is presented in the table while a larger value (24.55 MJm$^{-3}$ = 13.1 meV/cell) [27] at lower temperature has also been reported. Notice in Table 2 that for these three systems the calculations are in very reasonable agreement with experiments thus providing credibility to our predicted value for SmCoNiFe$_3$ of 4.93 meV/cell (9.21 MJm$^{-3}$). There is some sensitivity to this value with respect to nickel content and replacing 5% Co with 5% Ni lowers the MAE by 12%, see Table 2. The MAE for our proposed permanent magnet is thus smaller than SmCo$_5$ but still about twice that of Nd$_2$Fe$_{14}$B (4.9 MJm$^{-3}$) [10].

In regards to the computational details, all formation energies are obtained from the total-energy difference per unit cell between the specific compound or alloy and its components in their ground state. The total energy is calculated in the framework of density-functional theory with the generalized gradient approximation for the electron exchange and correlation interactions. For the ordered compounds (Fig. 1) we have used a full-potential linear muffin-tin orbitals (FPLMTO) method [28] while for the solid-solution alloys (inset of Fig. 1) we employ alloy theory within the so-called coherent potential approximation (CPA) in the exact muffin-tin orbitals (EMTO) method [29].

For the most advanced treatment of the electron correlations in SmCo$_5$ we are applying state-of-the-art dynamical mean-field theory where the Sm 4*f* electrons are assumed localized within the Hubbard I approximation that has proven to be appropriate for the rare-earth metals more generally [12]. The weakly correlated transition-metal 3*d* orbitals are tackled with the STPF impurity solver and the method is referred to as DMFT-HIA. The technical scheme and parameters are the same as those presented by Grånäs *et al.* [11], who also described the entire methodology and implementation more thoroughly. The DMFT is implemented in the FPLMTO [28] where the crystal is divided into muffin-tin spheres surrounding each atom and an interstitial region. Basis functions, electron densities, and potentials are expanded in these spheres without any geometrical restriction. For the DMFT, the basis of the valence electrons is constructed from 4*s*, 4*p*, and 3*d* states for Co and from semi-core 5*s* and 5*p* together with valence 6*s*, 6*p*, 5*d*, and 4*f* states for Sm, assuming a triple basis (three energy parameters) for 6*s* and 6*p* and a double basis (two energy parameters) for the remaining states. The spin-orbit interaction is included. The correlated basis functions are chosen to be the muffin-tin orbitals introduced in Refs. [30-32]. The Brillouin zone integration over the 108 k points is carried out using Fermi-Dirac (FD) broadening corresponding to room temperature (300 K). The chosen Hubbard U is 8 and 2.5 eV for Sm 4*f* and Co 3*d* electrons, respectively. For the Sm 4*f* electrons, the Hund's J is calculated (0.98 eV) while for the Co 3*d* electrons it is chosen (0.9 eV) as before [11].

The EMTO method is performed using a Green's-function technique based on the improved screened Korringa-Kohn-Rostoker method, where the one-electron potential is represented by optimized overlapping muffin-tin (OOMT) spheres [29]. The one-electron states are calculated exactly for the OOMT potentials. The outputs include the self-consistent Green's function of the system and the complete, non-spherically symmetric charge density. From this, the total energy is calculated using the full charge-density technique. We treat the 6*s*, 5*p*, 5*d*, and 4*f* for samarium and 4*s* and 3*d* states for cobalt and iron as valence states and the corresponding Kohn-Sham orbitals are expanded in terms of *spdf* exact muffin-tin orbitals. Integration over the Brillouin zone is performed using the special k point technique with 784 k points in the irreducible wedge of the zone (IBZ).

For the MAE calculations, we apply the SRM for the 4$f$ states on samarium within the FPLMTO method [28]. The approach is analogous to that of Steinbeck *et al.* [33] who obtained a reasonable MAE for SmCo$_5$ (~ 7 meV/cell) and the method is referred to as the DFT-SRM in the main text. For Sm, we use a double basis set with two energy parameters for the semi-core 5$s$ and 5$p$ and valence, 6$s$, 6$p$, 5$d$, and 5$f$ states. Similarly, for the TM atom, we include double basis semi-core 3$s$, 3$p$, and valence 4$s$, 4$p$, 3$d$, and 4$f$ states. For SmCoNiFe$_3$ the TM 2c sites are assumed to be a mix between Co and Ni that we model within the virtual crystal approximation (VCA). The VCA simply replaces the Co-Ni alloy with an average Co-Ni component with a nuclear charge of 27.5 and with 16.5 valence electrons. The VCA has been applied successfully for modeling magnetic moments in Fe-Co-Ni alloys [16] and the MAE in (Fe$_{1-x}$Co$_x$)$_2$B [34]. Because the magnitudes of the orbital moments are very important for the MAE we include a parameter-free orbital polarization (OP) correction [35] for the $d$ orbitals in addition to spin-orbit coupling and it is essential for in Fe, Co, and Ni [16, 35, 36]. The MAE is also very sensitive to technical parameters such as (i) number of k points in the IBZ and (ii) the type of broadening associated with the energy eigenvalues. To ensure convergence with respect to (i) we sample between 20-60 thousand k points in the IBZ and the MAE is converged within the number of significant digits displayed in Table 2. For the energy eigenvalue broadening we apply a room-temperature FD function. For $\alpha$-Co and SmCo$_5$ we also analyze the often-used Gaussian broadening scheme with 5, 10, 15, and 20 mRy energy widths. The result from the smallest Gaussian broadening is converged and the same within 2% to that of the FD 300 K function. Our MAE calculation for $\alpha$-Co reproduces the earlier result by Trygg *et al.* [36] when the same 10 mRy Gaussian is chosen.

For the calculation of the Curie temperature a mapping of the exchange interactions onto the Heisenberg Hamiltonian is performed. The T$_c$'s are obtained from a mean-field expression [37] and a classical Monte Carlo (MC) simulated-annealing method coupled to a heat bath [38] as implemented in UppASD code [39]. To thermalize the system 10$^6$ MC steps are used for each temperature that is varied in a range from 0 to 1400 K in steps of 100 K with a finer mesh (1 K) around T$_c$. The obtained T$_c$ corresponds to a peak in the calculated magnetic susceptibility. The exchange interactions used to calculate T$_c$ are computed from the FPLMTO method [28]. The sampling of the k points includes about 17000 in the IBZ and the Sm 4$f$ orbitals are treated in the SRM model as spin polarized core states. Spin-orbit interaction is neglected because otherwise the mapping onto the Heisenberg model is not possible. Most other parameters of the calculations are the same as those detailed above for the MAE computations.

After a thoughtful consideration of the physics involved with correlations of the 4$f$ and 3$d$ electrons we realize that a sensible and practical scheme for calculating the MAE is that of a DFT model within the SRM for the Sm 4$f$ states and orbital polarization for the TM 3$d$ states. All magnetic moments agree very well between the DMFT-HIA, the DFT-SRM, and polarized-neutron diffraction data for SmCo$_5$. The calculated electronic spectra also compare relatively well with photoemission spectrum except at very low binding energies where the 4$f$ contribution becomes imperative and only the DMF-HIA is correct. The compromise we make with the DFT-SRM is that the 4$f$ contribution is ignored but it is important to acknowledge, however, that this assumption is the same for all SmTM$_5$ alloys we have studied and changes of magnetic properties due to modifications of the 3$d$ contribution (doping with iron and nickel) are therefore defensible.

In conclusion, we have shown that substituting most of the cobalt with iron in SmCo$_5$ and doping with a small amount of nickel results in a new permanent magnet that we believe is possible to manufacture. It is predicted to have exceptional magnetic properties; a giant maximum energy product, a strong magnetic anisotropy, and a very high Curie temperature.

A. L. thanks A. Ruban, O. Peil, L. Vitos and I. L. thanks P. Thunström for technical support. O. E. acknowledges the support from VR, eSSENCE, and the KAW foundation (grants 2012.0031 and 2013.0020). This work performed under the auspices of the U.S. DOE by LLNL under Contract DE-AC52-07NA27344. This research is supported by the Critical Materials Institute, an Energy Innovation Hub funded by the U.S. DOE, Office of Energy Efficiency and Renewable Energy, Advanced Manufacturing Office.

**Figures**

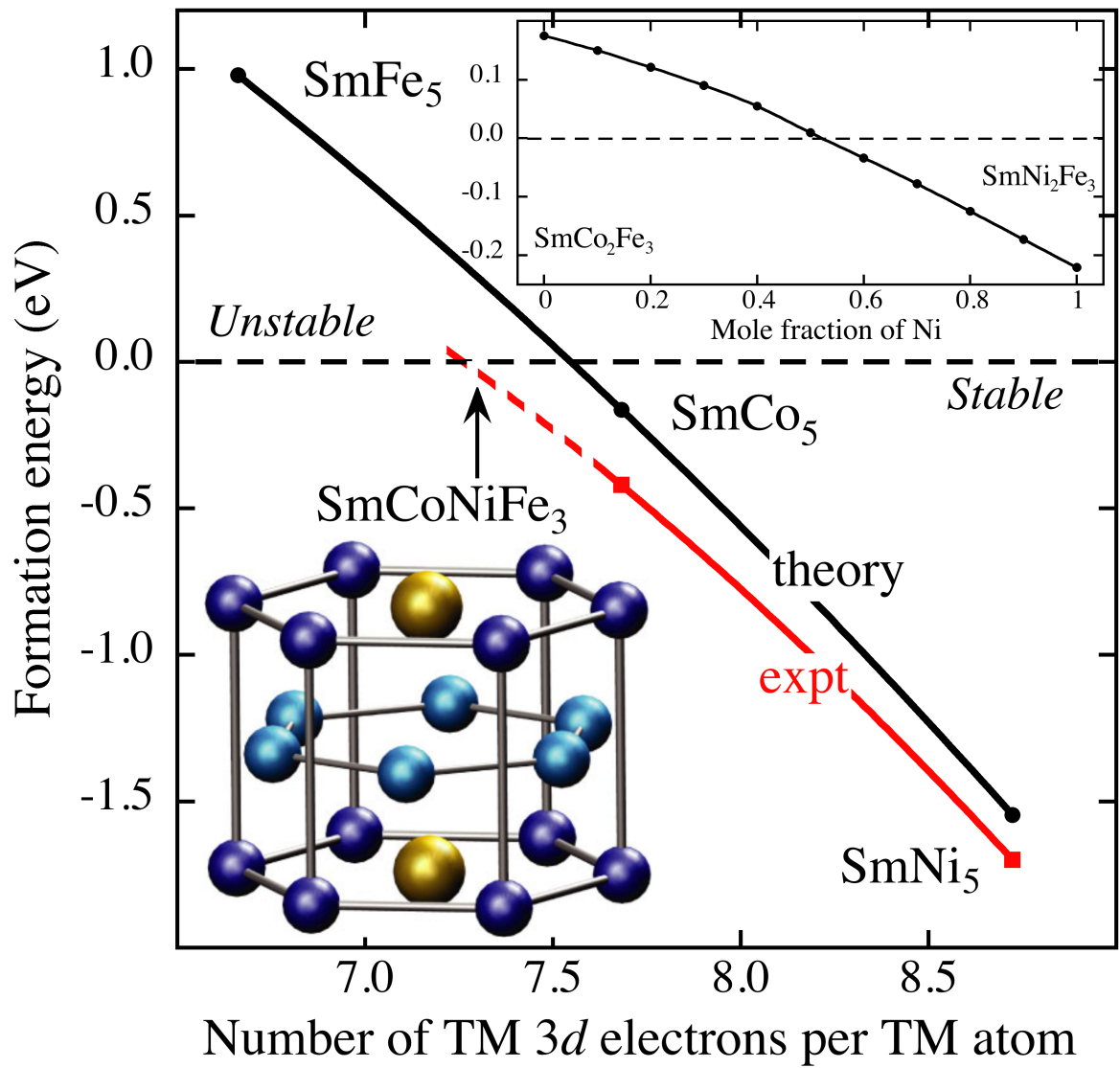

Fig. 1. Calculated (black circles) and experimental (red squares) [5, 6] formation energies (eV/cell). The red dashed line is an extrapolation. Negative energies indicate thermodynamic stability. The crystal-structure schematic is taken from Ref. [2].

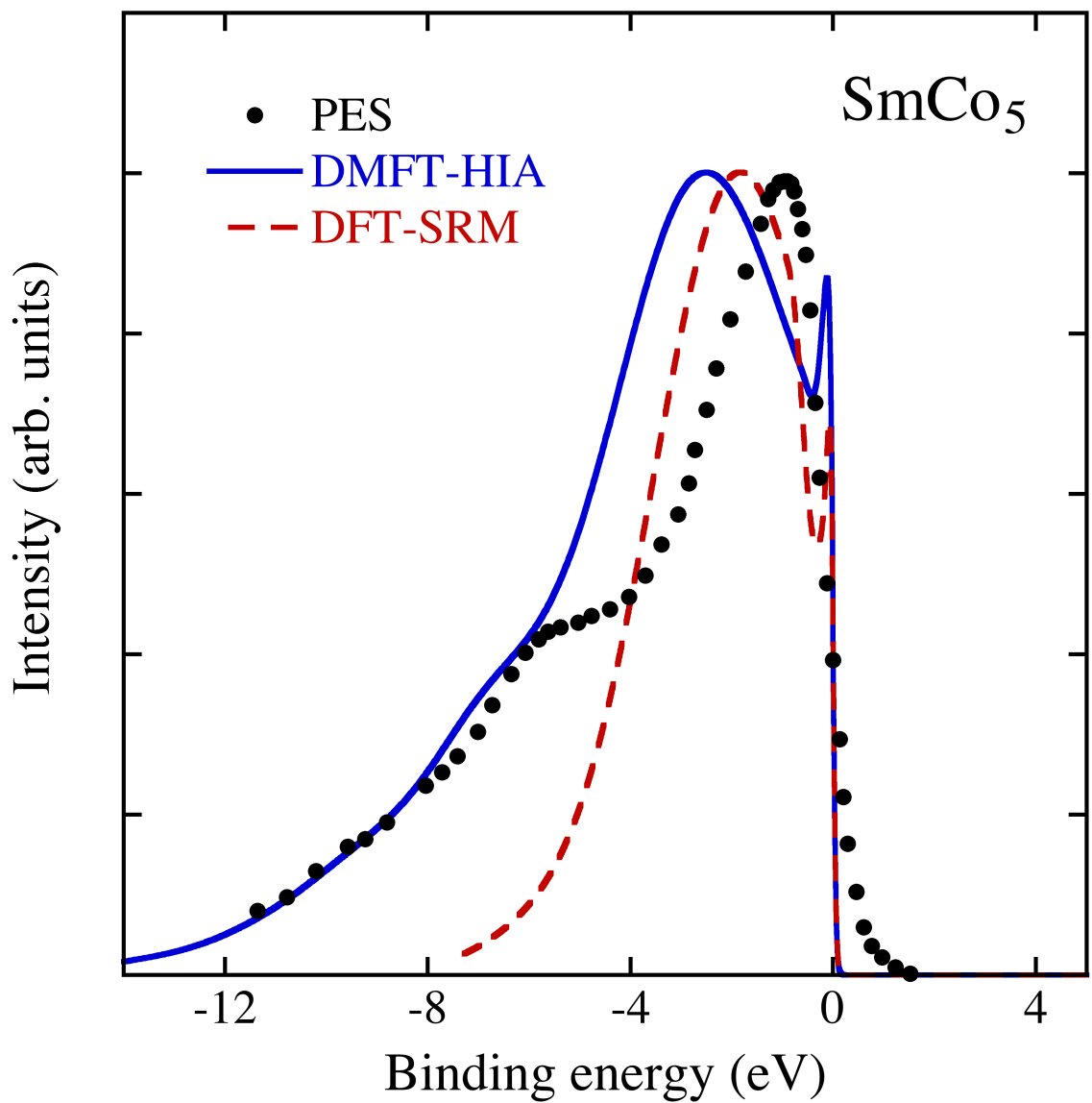

Fig. 2. The calculated photoemission spectra are obtained from total electronic density of states that are broadened due to a 300 K FD distribution, instrumental resolution, and lifetime of the photohole. The experimental photoemission spectrum (PES) is taken from Ref. [15].

**Tables**

| Method | Sm Total | Co$_1$ (2c) Spin, Orbital, Total | Co$_2$ (3g) Spin, Orbital, Total | Interstitial Spin | Total |
| --- | --- | --- | --- | --- | --- |
| DMFT-HIA | -0.06 | 1.54, 0.24, 1.78 | 1.52, 0.20, 1.72 | -0.46 | 8.20 |
| DFT-SRM | -0.30 | 1.61, 0.22, 1.83 | 1.60, 0.18, 1.78 | -0.43 | 8.27 |
| Polarized-neutron diffraction | -0.38 | –, –, 1.86 | –, –, 1.75 | – | 8.59 |

Table 1. Calculated and measured magnetic moments ($\mu_B$) for SmCo$_5$. The polarized-neutron diffraction results refer to low temperature (4.2 K) [17]. Other experimental reports for total moments for SmCo$_5$ are 7.8-8.9 $\mu_B$ given by Refs. [18, 19].

| Magnet | Theory | Experiment |
| --- | --- | --- |
| α-Co | 0.176 | 0.130 |
| SmCo$_5$ | 10.5 | 9.2 |
| SmCo$_2$Fe$_3$ | 7.55 | - |
| SmCoNiFe$_3$ | 4.93 | - |
| Sm(Co$_{0.45}$Ni$_{0.55}$)$_2$Fe$_3$ | 4.33 | - |
| SmNi$_5$ | 1.77 | 2.66 |

Table 2. Calculated and measured magnetic anisotropy energies (meV/cell). Experimental data are from Refs. [10, 25, 26].